\documentclass[a4paper,12pt]{article}

\usepackage{graphicx}
\usepackage{dcolumn}
\usepackage{epsfig}

\newcommand{\be}{\begin{equation}}
\newcommand{\ee}{\end{equation}}
\newcommand{\bd}{\begin{displaymath}}
\newcommand{\ed}{\end{displaymath}}
\newcommand{\BE}{\begin{eqnarray}}
\newcommand{\EE}{\end{eqnarray}}

\begin{document}


\title{Observed choices and underlying opportunities}
\author{Silvio Franz\thanks{
Centre scientifique d'Orsay, Universit\'e Paris-Sud 11, France },
Matteo Marsili\thanks{The Abdus Salam International Centre for Theoretical Physics, Strada Costiera 11, 34014 Trieste, Italy}, and
Paolo Pin\thanks{Dipartimento di Economia Politica, Universit\`a degli Studi di Siena, Piazza San Francesco 7, 53100 Siena, Italy}
}

\date{\today}

\maketitle

\begin{abstract}
Our societies are heterogeneous in many dimensions such as census, education, religion, ethnic and cultural composition. The links
  between individuals -- e.g. by friendship, marriage or collaboration
  -- are not evenly distributed, but rather tend to be concentrated
  within the same group.  This phenomenon, called {\em imbreeding
    homophily}, has been related to either (social) preference for
  links with own--type individuals ({\em choice--based} homophily) or
  to the prevalence of individuals of her same type in the choice set
  of an individual ({\em opportunity--based} homophily).
  Choices determine the network of relations we observe whereas opportunities pertain to the composition of the (unobservable) social network individuals are embedded in and out of which their network of relations is drawn.
  In this view, we propose a method that, in the presence of multiple data, allows one to distinguish between opportunity and choice based homophily. The main intuition is that, with unbiased opportunities, the effect of choice--based homophily gets weaker and weaker as the size of the minority
  shrinks, because individuals of the minority rarely meet and have the chance to
  establish links together. The occurrence of homophily in the limit of very small minorities is therefore an indicator of opportunity bias.
  We test this idea across the dimensions of race and
  education on data on US marriages, and across race on friendships in
  US schools.
\end{abstract}

\\



Integration is a major concern of our societies, whose relevance has
increased as an effect of globalization. The prevalence of relations
between individuals of the same type or community over links across
types -- a well known phenomenon called (inbreeding) \emph{homophily}
in sociology
\cite{LazarsfeldMerton54,Blau77,Marsden81,RytinaMorgan82,McphersonSmithlovin87,Marsden88,Mcpherson_alii01,Moody01} --
has been related to either {\em opportunity--based} or
\emph{choice--based} homophily \cite{McphersonSmithlovin87,Mcpherson_alii01}:
while the former (also called \emph{induced} homophily) refers to a prevalence of same--type neighbors in the
underlying social network, the latter reflects a bias towards
same--type links in the {\em collective} choice of mutual
relations, among those possible in a given neighborhood of the social
network. The relation between choice behavior and the underlying social network is a complex one. Indeed, the latter is often inferred from choice behavior -- friendship, marriage, co-authorship among scientists \cite{Newman01} -- which is relatively accessible to empirical studies. Second, opportunities constrains choices to the extent that choice behavior can hardly be related to choices of the individual, but rather to the choices of the population as a whole.
For example, Refs.  \cite{Currarini_alii09,Currarini_alii10} shows that individual
choices influence in non--trivial ways the aggregate outcome and Ref.
\cite{JegoRoehner07} argues that biased mixing of a minority may be
due to homophily of both majority and minority individuals.

Also, the opportunities which an individual faces when choosing
whom to establish a relation with may well be shaped by past choices
of that individual and others. For example, T.C. Schelling has vividly
shown that even very weak preferences for homophilous relations in
residential choice, can lead to strong spacial segregation
\cite{Schelling71,VinkovicKirman06}.
Finally, there are many dimensions (ethnical, religion, education, age, census
etc.) which are likely to influence, to different degrees, the
formation of links between individuals, and these are correlated in
complex ways. Disentangling their effect is a non-trivial task
\cite{BisinTopa03,Bisin_alii04}.

On the other hand, providing quantitative indicators to disentangle opportunity-based homophily from choice-based homophily is an important issue if, following Sen \cite{Sen}, if one regards constraints in the opportunities -- or freedom -- of individuals as a limiting factor for development.
This is particularly true in cases where the pattern of interaction is
shaped by institutions. For example, friendships between school
students is a matter of individual choice but their pattern of
interaction is largely shaped by institutions (clubs, sport teams,
academic tracking \cite{AcTrack},  etc). So while it is natural to
expect choice--based homophily, the presence of opportunity--based
homophily may be a matter of concern for policy makers.

Our aim, in the present work, is to show that the density dependence of standard homophily indicators can be used to disentangle the effects of opportunity--based homophily (OBH) and of choice--based homophily (CBH) for a minority group inside a larger population. The idea is that the network of relations that we observe in the data is a sub--network of the network of opportunities that all the individuals face, which is however usually not observable from the data.
In the absence of opportunity biases, this underlying unobserved network of opportunities is not biased and neutral to any minority.
In such unbiased network, CBH has an effect which is proportional to the size of the minority and, when the latter is small CBH is also negligible, simply because individuals of that minority have no opportunities to meet. Therefore, an excess of inter-type links in very small minorities must be due to OBH.

To be more precise, let $q$ be the ratio of same--type links for a
member of a minority and let $p$ be the relative size (frequency) of
this minority.  Following Coleman \cite{Coleman58}, Inbreeding
Homophily can be measured by the index
\begin{equation} \label{H}
H \equiv \frac{q  - p}{1-p} \ ,
\end{equation}
which is the excess of minority type links normalized so that baseline
homophily ($q=p$) \cite{Blau77} corresponds to $H=0$ and complete
segregation ($q=1$) yields $H=1$. \cite{wright}
The index $H$ can be expected to depend on social and cultural traits.
We assume that in similar cultural and social environments, with
different minority fractions $p$,  $H$ is a
well defined function of $p$ that reflects some of these
traits. In particular we associate to OBH the behavior of $H(p)$ for
small $p$. $H(p)$ depends non--linearly on $p$
\cite{Marsden81,Moody01,Currarini_alii07}, but for small $p$, we can
assume \cite{non-linear}
\begin{equation} \label{appr_H}
H (p) \simeq A + B p \ .
\end{equation}
The above observation on opportunities implies that we should have $A=0$ in a
population of homophilous individuals, with no OBH. Therefore, $A$ can
be taken as an index of OBH. This general observation can be detailed
in a simple probabilistic model, which explicitly takes into account
the two effects (see Model). OBH is modeled by the
frequency $\pi$ of the minority in the typical neighborhood of the
social network of a minority individual, for $p\to 0$. The actual
social relations are chosen on the social network thus defined, with
a same--type link of minority individuals being chosen $x$ times more
likely than a different--type link. $\pi$ and
$x$, in the model, can be derived from $A$ and $B$.

\begin{figure}[h]
\begin{minipage}{8cm}
\begin{center}
\includegraphics[width=8cm]{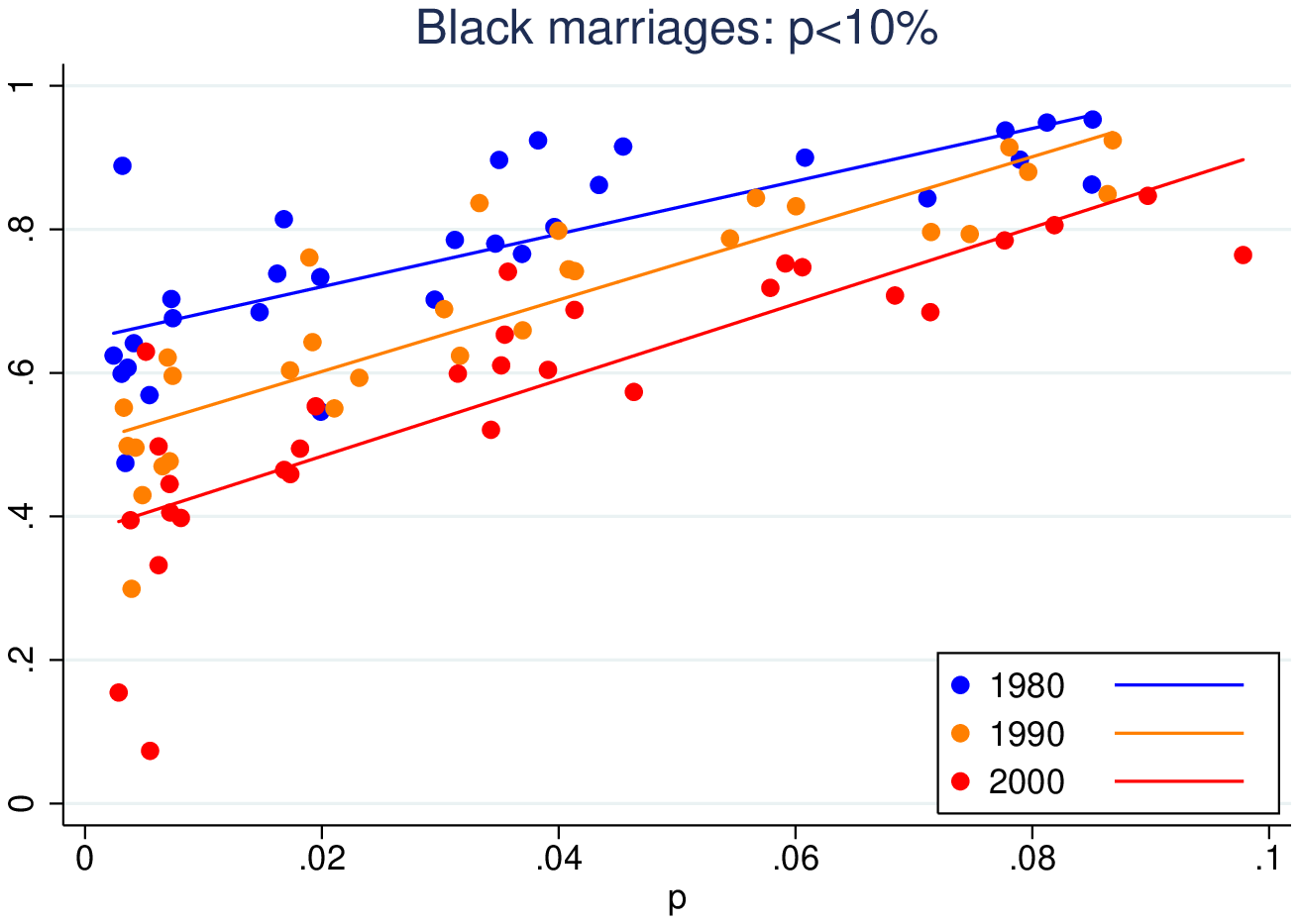}
\end{center}
\end{minipage}
\begin{minipage}{8cm}
\begin{center}
\includegraphics[width=8cm]{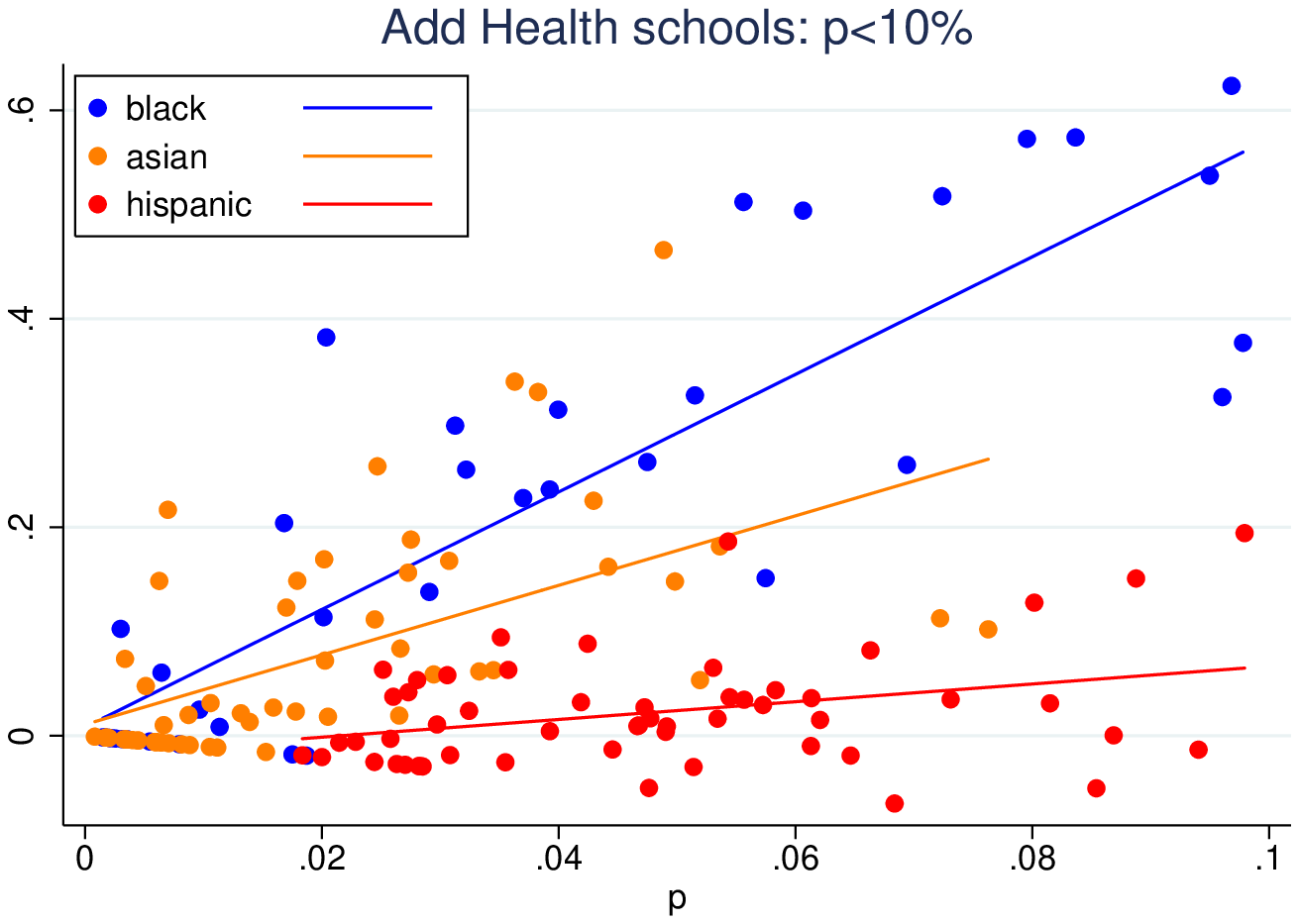}
\end{center}
\end{minipage}
\caption{IPUMS data for the Black minority (left) are based on same--race marriages in American States \cite{IPUMS}.
Each point represents the Black minority in one State in one of the three surveys (1980, 1990, 2000).
On the $x$--axis we have the percentage of the minority (the maximum threshold is $10\%$).
On the $y$--axis the Imbreeding Homophily measure defined in (\ref{H}).
Lines represent linear fits for each survey.
Add Health data (right) are based on same--race friendships in American schools \cite{AddHealth}.
Each point represents a minority in one school.
Linear fits are made for each minority.} \label{figure}
\end{figure}

We illustrate this idea on empirical data on marriages \cite{IPUMS}
and friendships \cite{AddHealth}, where individuals are identified by
race and (in marriages) by the level of education attained. The
datasets pertain to different environments -- single American States
for marriages \cite{IPUMS} or single schools for friendships
\cite{AddHealth} -- with different relative share $p$ of minority
individuals.
We make the strong assumption that every environment is otherwise identical.
For each of them, we measure the inbreeding homophily
index $H$ and compute $A$ and $B$ from a linear fit for small $p$,
$\pi$ and $x$ are computed from the model (see Model).  Fig.
\ref{figure} shows a sample of the results which are collected in
Table \ref{table}.
The left panel shows the fit of $H(p)$ for
marriages in the minority of Blacks, in three subsequent surveys
(1980, 1990 and 2000).
Points are largely scattered around the linear fit, indicating the presence of many factors in $H(p)$ which are not explained by density $p$. Still, there is a clear statistical trend.
First, we observe strong CBH ($x$), which is
also observed for other types (see Table \ref{table}). Also notice
that $A>0$, indicating OBH ($\pi>0$), though this effect has been
declining over time. OBH in marriages is observed also for other
minorities, but it is close to negligible for Native Americans, and its time
dependence is much weaker.
On the contrary, if we consider the minority of all those people having spent at least four years at college (the maximum score in the dataset, that we will call \emph{top education}),
CBH and OBH are remarkably stable across time with respect to this education--based classification, to the point that data
appears to lie on the same master line, though $p$ values have
increased on average of 46\% from 1980 to 2000.

Two considerations are necessary.
First, in all the three decades we considered all the married couples over the whole population, so our samples overlap. In order to test for this effect, we ran all the regression also for the sample of people under 40, in all the three decades.
Some of the temporal shift is thus anticipated, but the qualitative outcome is not affected.
As an example, the result for the Black minority is
$x=21.48 \pm 17.0$ and $\pi=0.060 \pm 0.03$ in 1980,
$x=13.21 \pm 10.7$ and $\pi=0.073 \pm 0.04$ in 1990,
$x=11.56 \pm 5.2$ and $\pi=0.026 \pm 0.01$ in 2000.
Second, our analysis singles out a single dimension (race or education), but these may be highly correlated among themselves and with other dimensions we do not observe, as discussed in Refs. \cite{Mcpherson_alii01}.
A projection on one dimension does not necessarily tell how strong its influence is, as the homophily may be due to a different characteristic, which is highly correlated with the one we are observing.
Cross correlations, within our analysis, can be probed by testing the hypothesis of independence in the choices and opportunity biases across two dimensions. This is done testing a version of our model based on multiple uncorrelated characteristics (see Model). In the case of the minority of top educated blacks, we fix the choice bias $x_{eb}$ that would be consistent with the hypothesis of independence and we estimate, under the same hypothesis, the parameters $\pi_e$ and $\pi_b$. This yields values (e.g. $\pi_e=0.483 \pm 0.41$ and $\pi_b=0.195 \pm 0.180$ in 2000) that are much larger than the original ones, thus rejecting the hypothesis.
This finding means, as expected, that the two dimensions are highly correlated, so that the top educated people in the matching sample of a top educated black are more likely to be black than if sampled at random from the population, and the other way round.

In school friendships (Fig. \ref{figure}, right panel), choice--based homophily is still
high and significative, but much less than for the marriages
considered above.
Also here $x$ strongly varies from race to race. 
It is moreover acceptable, for all the regressions and all the races,
to assume that $A$, and hence $\pi$ is equal to $0$, implying no OBH.
This is what we would assume from an environment like a school, where
the class formation should be independent on races \cite{AcTrack}.

Summarizing, we have proposed a method to disentangle choice--based
from opportunity--based sources of homophily (CBH and OBH
respectively). Our case study on two data sets shows that, for what
concerns marriages alone:
(i) OBH is stronger for \emph{top educated} people than for any racial
minority, but CBH is much weaker.
(ii) Looking at different time windows, for marriages, there is a clear decrease of both measures of homophily for Blacks between 1980, 1990 and 2000.
This time--dependence is not so evident for the other races and especially not for \emph{top educated} people.
For what concerns the racial dimension:
(iii) School friendships do not exhibit OBH (compared to the school population), while marriages do.
(iv) CBH is much stronger for marriages than for friendships.
(v) The values of both are strictly race--dependent:
Blacks exhibit the strongest CBH and (in marriages, if compared to the population of the American States) OBH;
Native and Hispanics exhibit the lowest values of both (which could be both accepted as uninfluent in the school data for Hispanics).

\begin{table} [h]
{
\tiny

\begin{tabular}{l|rrlrlrlrr}

{\bf IPUM marriages} &            &     {\bf } &            &     {\bf } &            &     {\bf } &     {\bf } &     {\bf } &     {\bf } \\

{\bf no threshold} & {\bf obs.} &    {\bf B} & {\bf \ $ \pm 95 \ \% $} &    {\bf A} & {\bf \ $ \pm 95 \ \% $} &    {\bf x} & {\bf \ $ \pm 95 \ \% $} & {\bf \ $ \pi_0 $} & {\bf \ $ \pm 95 \ \% $} \\
\hline
Top Educ. 1980 &         51 &      0,863 &        0,5 &      0,214 &       0,05 &       1,40 &        0,8 &      0,102 &       0,04 \\

Top Educ. 1990 &         51 &      0,778 &        0,3 &      0,218 &       0,04 &       1,27 &        0,6 &      0,109 &       0,03 \\

Top Educ. 2000 &         51 &      1,079 &        0,3 &      0,197 &       0,41 &       1,67 &        1,8 &      0,084 &       0,15 \\
\hline
{\bf \ $ 10 \ \% $} &     {\bf } &     {\bf } &     {\bf } &     {\bf } &     {\bf } &     {\bf } &     {\bf } &     {\bf } &     {\bf } \\
\hline
Black 1980 &         31 &      3,808 &        2,1 &      0,644 &       0,10 &      30,05 &       24,3 &      0,055 &       0,03 \\

Black 1990 &         31 &      4,470 &        2,1 &      0,511 &       0,10 &      18,69 &       11,8 &      0,050 &       0,02 \\

Black 2000 &         31 &      6,762 &        2,2 &      0,322 &       0,10 &      14,71 &        6,3 &      0,029 &       0,01 \\
\hline
Asian 1980 &         50 &      7,156 &        4,6 &      0,417 &       0,06 &      21,05 &       14,2 &      0,031 &       0,02 \\

Asian 1990 &         50 &      4,710 &        3,2 &      0,440 &       0,07 &      15,02 &       11,0 &      0,047 &       0,03 \\

Asian 2000 &         49 &      4,812 &        3,1 &      0,499 &       0,07 &      19,17 &       13,4 &      0,047 &       0,03 \\
\hline
Native 1980 &         50 &      6,449 &        1,6 &      0,169 &       0,04 &       9,34 &        2,6 &      0,019 &       0,01 \\

Native 1990 &         50 &      4,485 &        1,6 &      0,126 &       0,04 &       5,87 &        2,1 &      0,021 &       0,01 \\

Native 2000 &         49 &      4,450 &        1,6 &      0,154 &       0,04 &       6,22 &        2,4 &      0,025 &       0,01 \\
\hline
Hispanic 1980 &         45 &      2,877 &        1,8 &      0,355 &       0,06 &       6,92 &        4,5 &      0,065 &       0,03 \\

Hispanic 1990 &         42 &      3,730 &        1,8 &      0,234 &       0,06 &       6,36 &        3,3 &      0,040 &       0,02 \\

Hispanic 2000 &         40 &      3,802 &        1,6 &      0,274 &       0,08 &       7,21 &        3,4 &      0,044 &       0,02 \\
\hline
{\bf Add Health schools} &     {\bf } &     {\bf } &            &     {\bf } &            &     {\bf } &     {\bf } &     {\bf } &     {\bf } \\
\hline
     Black &         39 &     5,6385 &        1,0 &     0,0084 &       0,39 &       5,73 &        4,7 &      0,001 &       0,06 \\

     Asian &         56 &     3,3358 &        1,3 &     0,0109 &       0,04 &       3,41 &        1,3 &      0,002 &       0,01 \\

  Hispanic &         55 &     0,8504 &        0,7 &    -0,0184 &       0,04 &       0,82 &        0,6 &     0,000 &       0,02 \\
\hline
\end{tabular}

}
\caption{Every line represents a minority in one survey.
  For the first three the minority \emph{Top Educ.} represents all
  those people who have spent at least 4 years in college.
  For the remaining lines the minority is represented by a race.
  $n$ is the number of observations.
  We compute $A$, $B$ and their $95\%$ confidence interval, with a
  linear regression of $p$ versus $H$.
  In the cases concerning education there is no threshold on $p$
  (the reason for this is that $p$ has almost doubled in every State
  between 1980 and 2000).
  For the remaining regressions we take only those $p$ below $10\%$
  (results are qualitatively robust to a change of this threshold).
  We compute $x$, $\pi$, and their relative $95\%$ confidence
  interval, with the model described in the Appendix.} \label{table}
\end{table}

There are several interesting extensions of our analysis to other
dimensions such as religion or census, or to co-authorship networks in scientific research \cite{Newman01}.  We found non-trivial density dependence of homophily also across other dimensions, such as occupation, in marriage data. These cannot be easily related to properties of the underlying social network, as the type of work individuals choose may depend on whom they are married to.
The outcome of our analysis needs to be critically evaluated, as our distinction between choice and external constraints is theoretical at best.
If anything, it may help in identifying those institutional
constraints which hamper fruitful exchanges between members
of our society.

%




\section{The Model} \label{math}

We imagine a society whose individuals are ex-ante divided in
different types, whose number $N$ is fixed and large. Let $p$ be the
fraction of a particular minority in the population. The local
environment of each individual is defined by an underlying social
network, with $K$ links for each individuals.
This network is unobserved in the data.
$K$ is supposed to
represent the number of {\em possible} links from which the actual
relations (marriage, friendship, etc) of a particular individual are
drawn. We assume that $K$ is (much) larger than the actual number $k$
of relations each individual establishes, but much smaller than $N$
(for schools, $k\sim 6$ whereas $K\sim 30$ may be taken as the typical
class size, and $N$ is in the order of hundreds). Individuals are
distributed inhomogeneously on the social network, in such a way that
the average frequency of the minority in the neighborhood of a
minority individual is
\begin{equation} \label{segreg}
\bar{p} (p) = \pi + (1-\pi) p \ \  ,
\end{equation}
with $\pi \in (0,1)$. The relation is taken to be linear for
simplicity, with $\bar p(1)=1$.

We assume each individual of the minority has $k$ links, and we assign
them in the following way: {\em i)} choose an individual of the
minority at random, {\em ii)} if she still has an unassigned link,
choose one of the unassigned links in her neighborhood with a
statistical weight $1+x$ times larger for links to minority
individuals than to majority ones; {\em iii)} stop when all links of
the minority are assigned.
For marriages we consider a bipartite network in which all neighbors of an individual are of the opposite sex.
$x$ has a na\"ive interpretation in term of
utility in discrete choice models \cite{Mcfadden74}, but it also
reflects more complex aspects of the matching problem (see
e.g. \cite{Currarini_alii07}).

On average, each individual will have

\begin{equation} \label{fixed}
\left\{ \begin{array}{cl}
k \frac{ \bar{p} (1+x) }{(1+x) \bar{p} + (1 - \bar{p})} & \mbox{same--type links,}  \\
k \frac{ 1-\bar{p}}{(1+x) \bar{p} + (1 - \bar{p})} & \mbox{different--type links.}
\end{array} \right.
\end{equation}
Therefore the ratio $q$ of same--type links, in the whole population, over all links is $q \simeq
\frac{\bar{p} (1+x)}{1 + x \bar{p}}$, and this for small $p$ leads to

\begin{equation}
H(p)= \pi \frac{1+x}{1+ \pi x}+x \left( \frac{1-\pi}{1+\pi x}
\right)^2 p+ {\cal O}(p^2)
\end{equation}
from which we can read the values of $A$ and $B$ in
Eq. (\ref{appr_H}). Likewise, from $A$ and $B$ we can infer that
\begin{equation} \label{AB}
x \ \simeq \ \frac{B}{(1-A)^2} \ \mbox{ , } \ \pi \ \simeq \
\frac{A(1-A)}{1-A+B}  \ \ .
\end{equation}

As a check, we generated synthetic data sets using the model, with $x$
and $\pi$ given in Table \ref{table}, and performed a linear
regression of $H(p)$. The resulting values of $A'$ and $B'$ were found
to be within the 95\% confidence intervals reported in Table
\ref{table} for $A$ and $B$, in almost all cases. We attribute the
discrepancy to a systematic bias due to non--linear terms
in $H(p)$, as discussed above, which is particularly
strong when $x$ is large. These issues would require a more
sophisticated estimation techniques, which goes beyond the scope of the
present paper.

Finally, we considered a version of the model with two dimensions of classification, $a$ and $b$, for which we have separately identified $\pi_a$, $\pi_b$, $x_a$ and $x_b$ from the previous model.
Under the hypothesis of no correlation between the two characteristics,  individuals in the social network are either drawn of type $a$ or/and $b$, independently with probability $\pi_a$ and $\pi_b$, or they are drawn at random from the population. The analog of Eq. (\ref{segreg}) then reads
\[
\bar{p}_{ab} 
= \pi_a \pi_b + \pi_a (1-\pi_b) p_{b|a} + (1-\pi_a) \pi_b p_{a|b} + (1-\pi_a) (1-\pi_b) p_{ab}
\]
where $p_{ab}$ is the fraction of $ab$ individuals and $p_{a|b}$ is the fraction of agents of type $a$ among those of type $b$.
In the same  way we suppose that the choices' probabilities are uncorrelated, so that $x_{ab} = \frac{(1+x_a)(1+x_b)}{1 +(1+x_a)+(1+x_b)} - 1 = \frac{x_a x_b - 2}{3+x_a+x_b}$. In the statistical test, we compute $x_{ab}$ from the estimated $\hat x_a$ $\hat x_b$ and estimate $\pi_a$ and $\pi_b$ through a non-linear fit of $\bar{p}_{ab}$ on the data (errors are given by the robust variance estimation).
If the resulting values of $\pi_a$, $\pi_b$ are outside the 95\% confidence interval of the original values, we can discard the hypothesis of independence,
as we did above for the case of top educated blacks.












\begin{thebibliography}{99}
\bibitem{wright}
{In population genetics a
  similar index, known as F-statistics \cite{Wright22}, is used to
  quantify the deviations from the Hardy-Weinberg proportion between
  alleles.}
\bibitem{non-linear}
{%
  Both our probabilistic model and the real data have a non--linear
  behavior of $H$ for a larger range of $p$.  In some cases we found
  data to be better described by a non-linear relation
  $H(p)=\alpha+\beta p^\gamma+\ldots$ with $\gamma\approx 0.5$.  We
  tested however all our measures also with a non-linear error
  minimizing fit (based on the Model) and the qualitative nature of
  the conclusions discussed here does not change.}
\end{thebibliography}
\end{document}